\documentclass[showpacs,preprintnumbers,amsmath,amssymb,prd]{revtex4}

\usepackage{graphicx}
\usepackage{dcolumn}
\usepackage{bm}
\usepackage{natbib,hyperref,ifthen}

\newcommand{\ihmpc}{\, h\, {\rm Mpc}^{-1}}

\newcommand{\lyaf}{Ly$\alpha$ forest}

\newcommand{\vq}{\mathbf{q}}
\newcommand{\vx}{\mathbf{x}}
\newcommand{\vy}{\mathbf{y}}
\newcommand{\vk}{\mathbf{k}}

\newcommand{\fnl}{f_{\rm NL}}
\newcommand{\gnl}{g_{\rm NL}}
\newcommand{\orderfour}{\mathcal{O}\left(\delta_1^4\right)}

\begin{document}

\title{Primordial non-Gaussianity:  large-scale structure signature in the 
perturbative bias model}

\author{Patrick McDonald}
\email{pmcdonal@cita.utoronto.ca}
\affiliation{Canadian Institute for Theoretical Astrophysics, University of
Toronto, Toronto, ON M5S 3H8, Canada}

\date{\today}

\begin{abstract}

I compute the effect on the power spectrum of tracers of the large-scale 
mass-density field (e.g., galaxies) of primordial non-Gaussianity of the form 
$\Phi=\phi+\fnl\left(\phi^2-\left<\phi^2\right>\right)+\gnl\phi^3+...$, where 
$\Phi$ is proportional to the initial potential fluctuations and $\phi$ is a 
Gaussian field, using beyond-linear-order perturbation theory. I find that the 
need to eliminate large higher-order corrections necessitates the addition of a
new term to the bias model, proportional to $\phi$, i.e., $\delta_g=b_\delta
\delta+b_\phi\fnl\phi+...$, with all the consequences this implies for 
clustering statistics, e.g., $P_{gg}\left(k\right)=b_\delta^2 P_{\delta\delta}
\left(k\right)+2 b_\delta b_\phi\fnl P_{\phi\delta}\left(k\right)+b_\phi^2
\fnl^2 P_{\phi\phi}\left(k\right)+...~$.  This result is consistent with 
calculations based on a model for dark matter halo clustering, showing that the
form is quite general, not requiring assumptions about peaks, or the formation 
or existence of halos. The halo model plays the same role it does in the usual 
bias picture, giving a prediction for $b_\phi$ for galaxies known to sit in a 
certain type of halo. Previous projections for future constraints based on this
effect have been very conservative -- there is enough volume at $z\lesssim 2$ 
to measure $\fnl$ to $\sim \pm 1$, with much more volume at higher $z$. As a 
prelude to the bias calculation, I point out that the beyond-linear (in $\phi$)
corrections to the power spectrum of mass-density perturbations are naively 
infinite, so it is dangerous to assume they are negligible; however, the 
infinite part can be removed by a renormalization of the fluctuation amplitude,
with the residual $k$-dependent corrections negligible for models allowed by 
current constraints.

\end{abstract}


\maketitle

\section{Introduction}

Inflation \cite{1979JETPL..30..682S,1981PhRvD..23..347G,1982PhLB..108..389L,
1982PhRvL..48.1220A} has been tested successfully mainly through 
measurements of the power spectrum of primordial density perturbations
\cite{2006JCAP...10..014S,2008arXiv0803.0547K}.
These perturbations are expected to be nearly Gaussian and nearly scale 
invariant \cite{1981JETPL..33..532M,1982PhLB..115..295H,1982PhRvL..49.1110G,
1982PhLB..117..175S,1983PhRvD..28..679B}.  Testing the Gaussianity of the 
perturbations with increasing accuracy will be a major goal of future work
aimed at distinguishing different models (see \cite{2004PhR...402..103B} for a 
review of non-Gaussianity from inflation).
The simplest models of inflation predict that non-Gaussianity will be 
undetectably small \cite{1991PhRvD..43.1005S,1993PhRvL..71.1124L,
1994ApJ...430..447G,1995ApJ...454..552M,
2003NuPhB.667..119A,2003JHEP...05..013M,2004PhR...402..103B}, but
multifield models \cite{1997PhRvD..56..535L,2002PhRvD..66j3506B,
2003PhRvD..67b3503L,2007PhRvD..76d3526B,2006PhRvD..73h3522R,
2008PhRvD..78b3513I}, models where non-Gaussianity is generated during 
reheating \cite{2004PhRvD..69b3505D,2008PhRvD..77b3505S} or 
preheating \cite{2005PhRvL..94p1301E,2006PhRvD..73j6012B,2007PhRvD..75h6004B,
2008PhRvL.100d1302C},  
bouncing/ekpyrotic/cyclic models \cite{2007JCAP...11..010C,2008PhRvD..77f3533L,
2008PhRvL.100q1302B,2008PhRvD..78b3506L}, or inflation models based on nonlocal
field theory
\cite{2007JCAP...07..017B,2008JCAP...06..030B} can predict levels of 
non-Gaussianity near the 
present detection limits ($\fnl \sim 100$, as defined below). 
While there have been some hints of non-Gaussianity in the cosmic microwave
background (CMB) \cite{2007arXiv0710.2371J,2008PhRvL.100r1301Y}, the general 
consensus seems to be that nothing convincingly
primordial has been detected \cite{2003ApJS..148..119K,2007JCAP...03..005C,
2008arXiv0803.0547K}. 

Recently, \cite{2008PhRvD..77l3514D,2008ApJ...677L..77M,2008arXiv0806.1046A} 
showed that there 
should be a distinctive signature of non-Gaussianity in the large-scale power 
spectrum of dark matter halos, observable as galaxies, in the local model for
non-Gaussianity with curvature perturbations proportional to
\begin{equation}
\Phi = \phi +  \fnl \left(\phi^2 -\left<\phi^2\right>\right) +\gnl \phi^3+...
\end{equation}
\cite{2001PhRvD..63f3002K,2004PhR...402..103B,2006PhRvD..74j3003S,
2006PhRvD..74l3519B},
where I include the 3rd order term which is necessary to compute the power 
spectrum to 4th order in the small, Gaussian, perturbations $\phi$.
This form is only a special case of non-Gaussianity, but it is simple, and 
calculations using it should point the way to generalizations.
The idea of \cite{2008PhRvD..77l3514D}, to use the power spectrum, was a 
departure from the previously standard approach of studying 
non-Gaussianity in large-scale structure (LSS) using the bispectrum 
\cite{2000MNRAS.313..141V,2001PhRvL..86.1434F,2002MNRAS.335..432V,
2004PhRvD..69j3513S,2005MNRAS.361..824G,
2005MNRAS.364..620G,2005JCAP...10..010B,2006PhRvD..74b3522S,
2007PhRvD..76h3004S}
or other higher-order statistics \cite{2005PASJ...57..709H,
2006ApJ...653...11H,2007MNRAS.376..343K,
2007ApJ...658..669S,2008JCAP...04..014L,2008MNRAS.385.1613H}.  It was somewhat
surprising that this signal would be detectable, because the nonlinear 
corrections to the power spectrum in this 
model were generally assumed to be negligibly small (because fluctuations in 
$\phi$ are of order $10^{-5}$). Reference
\cite{2008JCAP...08..031S} verified and further explored this idea and applied 
it to real galaxy and quasar data sets.  They found observational constraints
$-1~(-23)<\fnl<+70~(+86)$, at 95\% (99.7\%) 
confidence, by combining LSS and CMB data.
The projected rms error from the Planck satellite, measuring the CMB 
anisotropy, is $\sigma_{\fnl} \sim 5$ \cite{2001PhRvD..63f3002K,
2008PhRvD..77l3006C}.

The calculations of
\cite{2008PhRvD..77l3514D,2008ApJ...677L..77M,2008JCAP...08..031S,
2008arXiv0806.1046A} were all
based on models for the clustering of dark matter halos.
The purpose of this paper is to investigate the result using a different 
approach -- the renormalized perturbative bias model of
\cite{2006PhRvD..74j3512M}.  
Instead of focusing on dark matter halos, this 
approach starts by assuming that the tracer density field is a 
completely general, unknown,
function of the local mass-density field and Taylor expanding this 
function in the mass-density fluctuations, leading to the 
form \cite{1993ApJ...413..447F}
\begin{equation}
\delta_g\left(\delta\right)=c_1 \delta + 
\frac{1}{2}c_2 \left(\delta^2-\sigma^2\right)+
\frac{1}{6}c_3 \delta^3+\epsilon+\orderfour~,
\end{equation}
where $\delta$ is the mass-density perturbation, 
$\sigma^2=\left<\delta^2\right>$, and
$\delta_g$ is the tracer density perturbation (I will usually refer to 
the tracer as galaxies, although there are many other possibilities like
quasars \cite{2008JCAP...08..031S},
the \lyaf\ \citep{2005ApJ...635..761M,2006ApJS..163...80M,2006MNRAS.365..231V},
galaxy cluster/Sunyaev-Zel'dovich effect measurements
\citep{2008RPPh...71f6902A}, and possibly
future 21cm surveys \citep{2005MNRAS.364..743N,2008PhRvL.100i1303C}).
The unknown coefficients
of the Taylor series have become the bias parameters $c_i$ and a random 
noise variable $\epsilon$ has been added to allow for stochasticity in the 
relation (e.g., from shot noise, at least).   
This type of bias model has been explored in many papers
\cite{1995ApJS..101....1M,
1998MNRAS.301..797H,
1998ApJ...504..607S,1999ApJ...510..541T,1999ApJ...522...46T,
1999ApJ...520...24D,2000ApJ...537...37T,2001ApJ...556..641H}.  
The new 
development in \cite{2006PhRvD..74j3512M} was to show that large
higher-order corrections in this perturbative approach can be eliminated by
redefining the bias parameters to absorb them, motivated by the way that masses
and
coupling constants are redefined to absorb divergent loop corrections in 
quantum field theory \cite{Peskin:1995ev}.  
I argued that this 
renormalization approach is an improvement over the previous approach 
of defining the galaxy density to be a function of a smoothed
mass-density field (so the higher-order terms in the Taylor series 
can be kept explicitly small \cite{1993ApJ...413..447F,1998MNRAS.301..797H}), 
because the 
smoothing needed is so extreme that it directly affects the scales of interest 
\cite{2007PhRvD..75f3512S,2008PhRvD..78b3523S},
and because, in the unrenormalized approach, the higher-order corrections 
modify the power 
on arbitrarily large, ideally truly linear, scales \cite{1998MNRAS.301..797H}.
Reference \cite{2008arXiv0805.2632J} found extremely good agreement between the 
renormalized model and the clustering of galaxies in numerical simulations. 
Generally, the value of perturbation theory (PT) for describing LSS will only
increase as observational measurements become more precise, because this 
increases the range 
of scales where corrections to linear theory are important but are still small 
enough
to be treated perturbatively.  The value of PT has been further enhanced 
recently by
the introduction of several renormalization methods applicable to the
mass-power spectrum calculation \cite{2006PhRvD..73f3519C,2006PhRvD..73f3520C,
2007PhRvD..75d3514M,2007JCAP...06..026M,
2007PhRvD..76h3517I,2008ApJ...674..617T,2008PhRvD..77b3533C,
2008MPLA...23...25M,2008PhRvD..77f3530M,
2008PhRvD..78h3503B}.

To complete the relation between primordial non-Gaussianity and final density, 
note that 
the initial density field is related to $\Phi$ through the transfer function
from primordial to late-time linear fluctuations, $T(k)$, and the 
Poisson equation
\begin{equation}
\delta_1\left(\vk,a\right) = 
\frac{2}{3}\frac{c^2 D\left(a\right)}{\Omega_{m,0}H_0^2 } k^2 
T\left(k\right)\Phi_\vk \equiv M\left(k,a\right) \Phi_\vk ~,
\label{eqPoisson}
\end{equation}
using the definitions of \cite{2007PhRvD..76h3004S},
where I will use the subscript on $\delta$ to indicate the order in the 
initial density perturbations, not the Gaussian field $\phi$ (I will use
$\delta_L$ to indicate the fully linear, including in $\phi$, density 
perturbation).  The growth factor is normalized so $D\left(a\right)=a$ in the 
matter dominated era.  
The transfer function is time independent and normalized by
$T\left(k\rightarrow 0\right)\rightarrow 1$.  These definitions make 
$\Phi_\vk$ time independent and close to scale invariant.
The perturbative density field is written as a Taylor series in $\delta_1$, 
i.e., $\delta=\delta_1+\delta_2+\delta_3+...$, with 
$\delta_i \sim \mathcal{O}\left(\delta_1^i\right)$, e.g.,
\begin{equation}
\delta_2\left(\vk\right)= 
\int \frac{d^3\vq}{\left(2 \pi\right)^3}\delta_1\left(\vq\right)
\delta_1\left(\vk-\vq\right) J_S^{\left(2\right)}\left(\vq,\vk-\vq\right)~.
\end{equation}
where
\begin{equation}
J_S^{(2)}(\vk_1,\vk_2)=
\frac{5}{7}+\frac{1}{2}\frac{\vk_1\cdot\vk_2}{k_1 k_2}
\left(\frac{k_1}{k_2}+\frac{k_2}{k_1}\right)+\frac{2}{7}
\left(\frac{\vk_1\cdot\vk_2}{k_1 k_2}\right)^2
\end{equation}
is given by standard LSS
perturbation theory (see \cite{2002PhR...367....1B} for a review).

The rest of the paper goes as follows:
To put the perturbative bias calculation on a firm foundation, in 
Sec. \ref{secmassdensity}, I compute the mass power spectrum in the 
non-Gaussian model.
In Sec. \ref{secbias}, I calculate the power spectrum of a biased tracer. 
Finally, in Sec. \ref{secconclusions}, I discuss the results. 

\section{Mass-density power spectrum \label{secmassdensity}}

In  this section I compute the power spectrum of mass-density fluctuations in 
the non-Gaussian model.  I note in advance that infrared
divergences will appear, i.e., sensitivity to arbitrarily large scales, which 
suggests that a more sophisticated calculation 
than the usual, essentially Newtonian, LSS perturbation theory might be 
required.
The point of this section is simply to show that the usual LSS PT
can in fact be 
pursued self-consistently, without ignoring any divergences, imposing arbitrary
cutoffs, or integrating into
a regime where the calculations are not valid.  All of the 
sensitivity to very large (and very small) scales can be absorbed into the 
overall normalization of the
power spectrum, so there is no need to worry about exactly where 
a physical large-scale cutoff might appear, as long as it is much larger than 
the scale of the observations.  For more
physical discussions of infrared divergences in inflation, see 
\cite{1987PhLB..197...66A,2004PhRvD..70d3533B,2006PhRvD..73b1301B,
2006JCAP...06..015L,2007JCAP...12..016L,2008JCAP...04..030R,
2008JCAP...01..015B,2008JCAP...04..025E}.
Note that there are no divergences in standard LSS PT calculations of the mass 
power spectrum starting from Gaussian primordial perturbations, for a typical
$\Lambda$CDM power spectrum (divergences for pure power law linear 
power spectra
are discussed in \cite{1994ApJ...431..495J,1996ApJ...456...43J,
1996ApJ...473..620S}).

The power spectrum to 4th order in $\delta_1$ is 
\begin{equation}
P_{\delta\delta}\left(k\right)=
P_{\delta_1 \delta_1}\left(k\right)+2~P_{\delta_1 \delta_2}\left(k\right)+
2~P_{\delta_1 \delta_3}\left(k\right)+
P_{\delta_2 \delta_2}\left(k\right)+...
\end{equation}
The most interesting term turns out to be the first one, 
\begin{equation}
P_{\delta_1\delta_1}\left(k\right) = M^2\left(k,a \right)  
P_{\Phi\Phi}\left(k\right).
\end{equation}
with
\begin{equation}
P_{\Phi \Phi}\left(k\right) = P_{\phi \phi}\left(k\right)
+2~\gnl P_{\phi \phi^3}
+ \fnl^2 P_{\phi^2 \phi^2}\left(k\right)+...
\label{eqPPhiPhi}
\end{equation}
The first term here is simply the usual lowest order power spectrum, which I
assume is given by $P_{\phi \phi}\left(k\right)=A k^{n_s-4}$.  
The 2nd term is divergent, in a very simple way:
\begin{equation}
2~\gnl P_{\phi \phi^3}\left(k\right) = 
6~\gnl~\sigma^2_{\phi \phi} P_{\phi \phi}\left(k\right)
\end{equation}
where
\begin{equation}
\sigma^2_{\phi \phi} = \int \frac{d^3 \vq}{\left(2\pi\right)^3}
P_{\phi \phi}\left(q\right) ~.
\end{equation}
For $n_s<1$, $\sigma^2_{\phi \phi}= A 
\left(\epsilon^{n_s-1}-\Lambda^{n_s-1}\right)/
2 \pi^2 \left(1-n_s\right)$ is infrared divergent, where $\epsilon$ is the
minimum $q$ and $\Lambda$ is the maximum $q$ (which we would like to take to 
zero and infinity, respectively).  The solution to this divergence is obvious:
the divergent term just renormalizes the amplitude of the power spectrum, 
i.e., $A\rightarrow A+6~\gnl~\sigma^2_{\phi \phi}$.  

The third term in Eq. (\ref{eqPPhiPhi}) is slightly more subtle:
\begin{equation}
\fnl^2 P_{\phi^2 \phi^2}\left(k\right) = 
2 \fnl^2 \int \frac{d^3 \vq}{\left(2\pi\right)^3}
P_{\phi \phi}\left(q\right) P_{\phi \phi}\left(\left|\vk-\vq\right|\right) 
~\stackrel{\epsilon/k \rightarrow 0}{\longrightarrow}~
\left[4 \fnl^2 \sigma^2_{\phi \phi}\right] P_{\phi \phi}\left(k\right) ~.
\end{equation} 
There is again an infrared divergence for $n_s<1$, but it does not look like a 
simple $k$-independent renormalization of the amplitude, because,
after we absorb the divergent part, there will be a $k$-dependent 
piece left over, i.e., 
\begin{equation}
\fnl^2 P_{\phi^2 \phi^2}\left(k\right) = 
\left[4 \fnl^2 \sigma^2_{\phi \phi}\right] 
P_{\phi \phi}\left(k\right)+
2 \fnl^2 \int \frac{d^3 \vq}{\left(2\pi\right)^3}\left[
P_{\phi \phi}\left(q\right) P_{\phi \phi}\left(\left|\vk-\vq\right|\right)
-P_{\phi \phi}\left(q\right) P_{\phi \phi}\left(k\right)
-P_{\phi \phi}\left(k\right) P_{\phi \phi}\left(\left|\vk-\vq\right|\right)
\right]~,
\end{equation} 
where I have written the two equivalent negative terms in the integral to 
emphasize that half of the divergence comes from $q\rightarrow 0$ and half 
from $q\rightarrow k$.

Subtracting the divergent part has now introduced a new source of trouble, 
unfortunately, in that the 
subtracted terms very nearly diverge as $q\rightarrow \infty$,
e.g., for $n_s=0.96$ a cutoff $\Lambda \gtrsim 10^{50} k$ is 
required for convergence to 1\% accuracy.  
The solution is to pull out another factor renormalizing the 
amplitude, as follows:
\begin{eqnarray}
\fnl^2 P_{\phi^2 \phi^2}\left(k\right) &=&
P_{\phi \phi}\left(k\right)\left(
4 \fnl^2 \sigma^2_{\phi \phi} + 2\fnl^2
\int \frac{d^3 \vq}{\left(2\pi\right)^3}\left[
\frac{P_{\phi \phi}\left(q\right) 
P_{\phi \phi}\left(\left|\vk_\star-\vq\right|\right)}
{P_{\phi \phi}\left(k_\star\right)}
-P_{\phi \phi}\left(q\right) 
-P_{\phi \phi}\left(\left|\vk_\star-\vq\right|\right)
\right]\right) \\ \nonumber
&+&
2 \fnl^2 P_{\phi \phi}\left(k\right) 
\int \frac{d^3 \vq}{\left(2\pi\right)^3}\left[
\frac{
P_{\phi \phi}\left(q\right) P_{\phi \phi}\left(\left|\vk-\vq\right|\right)}
{P_{\phi \phi}\left(k\right)} 
-P_{\phi \phi}\left(\left|\vk-\vq\right|\right)-
\frac{
P_{\phi \phi}\left(q\right) 
P_{\phi \phi}\left(\left|\vk_\star-\vq\right|\right)}
{P_{\phi \phi}\left(k_\star\right)} 
+P_{\phi \phi}\left(\left|\vk_\star-\vq\right|\right)
\right] ~.
\end{eqnarray} 
I have simply added and subtracted $P_{\phi \phi}\left(k\right)$ times some
constant piece, with the constant chosen to cancel the correction term at some
arbitrary scale $k_\star$.  Note that we could drop the terms in the integral
that are simply integrals over 
$-P_{\phi \phi}\left(\left|\vk-\vq\right|\right)$ and
$P_{\phi \phi}\left(\left|\vk_\star-\vq\right|\right)$, because they exactly
cancel.  This would leave us with the form we would have found by starting with
the second form of renormalization (involving $k_\star$) instead of going 
through 
the initial step of subtracting the truly infinite part, i.e., the first
renormalization was not strictly necessary.  I leave the 
equation in the more complicated looking form because the simpler one would 
involve
cancellation between pieces of the integral that diverge at different values
of $q$ (i.e., $k$ and $k_\star$), which makes the integral less straightforward
to evaluate.

The power spectrum of $\Phi$ is now simply
\begin{eqnarray}
P_{\Phi \Phi}\left(k\right) &=& P_{\phi \phi}\left(k\right) \\ \nonumber &+&
2 \fnl^2 P_{\phi \phi}\left(k\right)\int \frac{d^3 \vq}{\left(2\pi\right)^3}
\left[\frac{P_{\phi \phi}\left(q\right) 
P_{\phi \phi}\left(\left|\vk-\vq\right|\right)}
{P_{\phi \phi}\left(k\right)}-
P_{\phi \phi}\left(\left|\vk-\vq\right|\right)-
\frac{P_{\phi \phi}\left(q\right)
P_{\phi \phi}\left(\left|\vk_\star-\vq\right|\right)}
{P_{\phi \phi}\left(k_\star\right)}+
P_{\phi \phi}\left(\left|\vk_\star-\vq\right|\right)\right]
\end{eqnarray} 
where $P_{\phi \phi}$ now has the renormalized amplitude, essentially 
disconnected from the original amplitude.
Inserting $P_{\phi \phi}\left(k\right)=A k^{n_s-4}$, we find, at least to a 
very good approximation
\begin{equation}
P_{\Phi \Phi}\left(k\right) \simeq P_{\phi \phi}\left(k\right)+
\fnl^2 A^2 \frac{2}{\pi^2\left(n_s-1\right)}
\left(k^{2 n_s-5}-k^{n_s-4}k_\star^{n_s-1}\right)
= P_{\phi \phi}\left(k\right)\left(1+
\frac{4 \fnl^2 }{\left(n_s-1\right)}
\frac{A \left(k^{n_s-1}-k_\star^{n_s-1}\right)}{2 \pi^2}\right)~.
\end{equation}
Note that this result is well behaved in the scale-invariant limit,
$\left(n_s-1\right)\rightarrow 0$, when
$\left(k^{n_s-1}-k_\star^{n_s-1}\right)/\left(n_s-1\right)\rightarrow 
\ln\left(k/k_\star\right)$ (this would have been a disaster without the 
second renormalization, introducing $k_\star$, of the UV near-divergence, which 
becomes a true divergence in this limit).

Now we can estimate the change in apparent slope, 
\begin{equation}
\Delta n_s=d\ln 
\left(1+ \frac{4 \fnl^2 }{\left(n_s-1\right)}
\frac{A \left(k^{n_s-1}-k_\star^{n_s-1}\right)}{2 \pi^2}\right) /d\ln k
\simeq
4 \fnl^2 \Delta^2_{\phi \phi}\left(k\right)~,
\end{equation}
where $\Delta^2_{\phi \phi}\left(k\right)=A k^{n_s-1}/2 \pi^2$,
and change in the running of the slope
\begin{equation}
\Delta \alpha_s=\frac{d \Delta n_s}{d\ln k}\simeq
\left(n_s-1\right) \Delta n_s~.
\end{equation}
For a realistic model 
$\Delta^2_{\phi \phi}\left(k\right) \simeq 8\times 10^{-10}$ 
so the change in slope is quite small, e.g., for an unrealistically 
large $\fnl=1000$, 
$\Delta n_s\simeq 0.0033$, which is roughly the expected precision of a 
measurement of the slope using Planck data
\cite{2006PhRvD..73h3520T,2007PhRvD..76f3009M}. 
The bottom line of this calculation is that we can safely ignore the 
non-Gaussian contribution to $P_{\delta_1 \delta_1}$, for realistic models.
Note that, even if the new contribution to the slope or running had been 
significant, it would not necessarily have been distinguishable from a change
in the underlying inflation model, i.e., the bare slope or running.  It would
still need to be kept in mind, however, in order to correctly interpret 
measurements of the power spectrum as constraints on the inflation model.

The rest of the calculation of the mass power spectrum is less interesting, 
\begin{equation}
2 P_{\delta_1 \delta_2}\left(k\right) =
4 \fnl \int \frac{d^3 \vq}{\left(2\pi\right)^3} 
J^{\left(2\right)}\left(\vq,\vk-\vq\right) P_{\phi \delta_L}\left(q\right)
\left[2
P_{\phi \delta_L}\left(k\right) M\left(\left|\vk-\vq\right|\right)+
P_{\phi \delta_L}\left(\left|\vk-\vq\right|\right) M\left(k\right)\right]
\end{equation} 
where $P_{\phi \delta_L}(k)\equiv M\left(k\right) P_{\phi \phi}\left(k\right)$.
For realistic power spectra, $P_{\delta_1 \delta_2}(k)$ is comfortably 
convergent in both limits, and produces a subpercent effect for $\fnl<10^5$. 
The terms 4th order in $\delta_1$ are of course just the usual nonlinear 
correction to the Gaussian linear theory power spectrum.

To be clear, the conclusion of this section is that the non-Gaussianity of 
the initial 
conditions in the local model does not significantly affect the mass power 
spectrum, verifying what has been previously assumed.
Note that \cite{2007PhRvD..76h3517I} considers a different form of 
renormalized PT with non-Gaussianity, 
but the results are not directly comparable because their specific example is
a different form of non-Gaussianity (local in density).

\section{Bias \label{secbias}}

This section contains the primary calculation of this paper:  the power 
spectrum of biased tracers of the density field.
I start by assuming the bias takes the form of a local Taylor series in 
density
\begin{equation}
\delta_g\equiv \rho_g/\bar{\rho}_g-1=
c_\delta~\delta +
\frac{1}{2}~ c_{\delta^2}~\left( \delta^2-\sigma^2\right) +
\frac{1}{3!}~ c_{\delta^3}~ \delta^3 +\epsilon+\orderfour~.
\end{equation}
The first, and only, new nontrivial term related to non-Gaussianity is the 
linear-quadratic cross-term
\begin{eqnarray}
c_\delta c_{\delta^2} P_{\delta \delta^2}\left(k\right)&=&2 c_\delta 
c_{\delta^2} \fnl
\left[
M\left(k\right)
\int \frac{d^3 \vq}{\left(2\pi\right)^3} P_{\phi \delta_L}\left(q\right)
P_{\phi \delta_L}\left(\left|\vk-\vq\right|\right)
+
2 P_{\phi \delta_L}\left(k\right)
\int \frac{d^3 \vq}{\left(2\pi\right)^3} P_{\phi \delta_L}\left(q\right)
M\left(\left|\vk-\vq\right|\right) \right] \\ \nonumber &+&
\fnl{\rm -independent~ parts}~.
\label{eqnontrivialbias}
\end{eqnarray}
The second term in the brackets is not well behaved in the limit 
$q\rightarrow \infty$, where it becomes $\left[c_\delta c_{\delta^2} 4 \fnl
\sigma_{\delta\delta}^2\right] 
P_{\phi \delta_L}\left(k\right)$ (this term comes from
the part of $\delta\left(\vx\right)\delta^2\left(\vy \right)$ 
proportional to $\delta\left(\vx\right)\delta\left(\vy\right) 
\phi\left(\vy\right) \nabla^2\phi\left(\vy\right)$).
It nearly diverges, as discussed above, in the same sense
as the unsmoothed linear density variance nearly diverges, i.e., the 
cutoff-dependent limit goes like $\Lambda^{n_s-1}$ (at least until one
hits the cold dark matter free streaming scale, which actually happens before 
the integral would otherwise converge \cite{2005JCAP...08..003G}
-- the fact that that scale enters this discussion should make it clear why
something needs to be done about this term).  
It is not immediately obvious how to deal with this term.  Ignoring it is not 
a good option, because it can become large relative to the leading order term,
at small $k$, i.e.,  
\begin{equation}
\frac{c_\delta c_{\delta^2} 4 \fnl \sigma_{\delta\delta}^2
P_{\phi \delta_L}\left(k\right)}
{c_\delta^2  P_{\delta_L \delta_L}\left(k\right)} = 
\frac{c_{\delta^2}}{c_\delta}4 \fnl
\sigma_{\delta\delta}^2 M^{-1}\left(k\right) \sim  29
\left(\frac{\fnl}{100}\right)
\left(\frac{c_{\delta^2}}{c_\delta} \right)
\left(\frac{\sigma_{\delta\delta}^2}{100}
\right)\left(\frac{k}{0.01 \ihmpc}\right)^{-2}~.
\end{equation} 
Taking this result at face value is not an option, both because it is a large
correction in what is supposed to be a perturbative expansion, and because it
is ridiculous in LSS perturbation theory to trust the quantity 
$\sigma_{\delta\delta}^2$ which requires integration many orders of 
magnitude into the nonlinear regime for convergence.
We cannot remove this term by renormalizing the linear bias
parameter, because the $k$ dependence of the linear term, proportional to 
$P_{\delta_L \delta_L}\left(k\right)$, is not consistent with
the $k$ dependence of this term, proportional to 
$P_{\phi \delta_L}\left(k\right)$ (we
could always cancel the value of this term at one point $k_\star$, but the
differing $k$ dependence would mean that the term would become large 
again very quickly as we went away from $k_\star$).  

The solution seems to be to introduce a new term in the bias formula
$c_\phi \fnl \phi$, i.e.,
\begin{equation}
\delta_g\equiv \rho_g/\bar{\rho}_g-1=
c_\delta~\delta +c_\phi \fnl \phi+
\frac{1}{2}~ c_{\delta^2}~\left( \delta^2-\sigma^2\right) +
\frac{1}{3!}~ c_{\delta^3}~ \delta^3 +\epsilon+\orderfour~.
\end{equation}
Then the power spectrum will contain the term 
$2 c_\delta c_\phi \fnl P_{\phi \delta_L}\left(k\right)$, which is needed to 
absorb the divergence, i.e., the bare value of $c_\phi$ (which might be zero)
will be renormalized to $b_\phi = c_\phi + 
2 c_{\delta^2} \sigma_{\delta\delta}^2$. 

\begin{figure}
\resizebox{\textwidth}{!}{\includegraphics{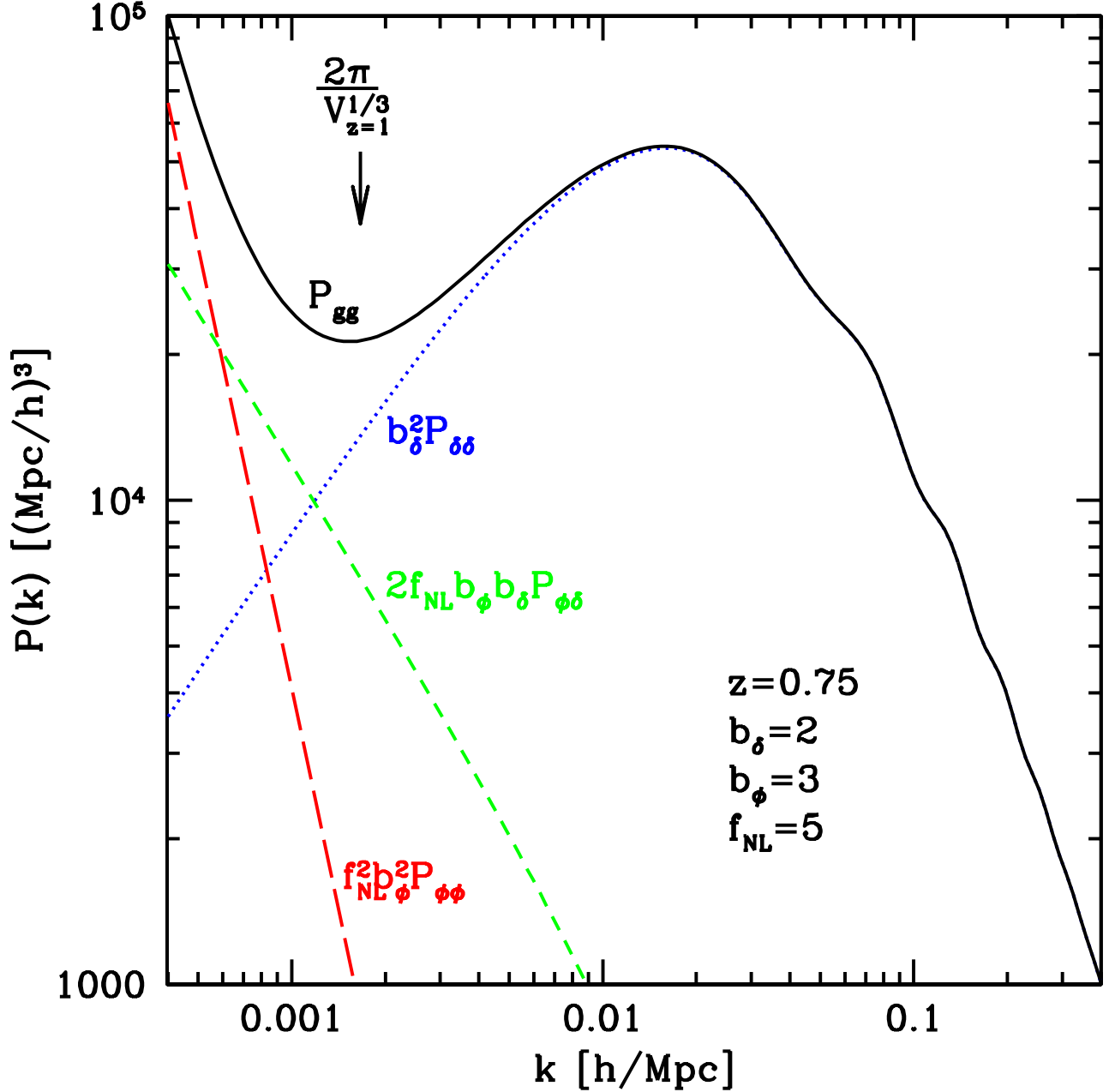}}
\caption{
Real-space power spectrum for $\fnl=5$, at $z=0.75$, of galaxies with 
$b_\delta=2$, $b_\phi=3$ (black, solid line).  Also shown are the components
of the power spectrum, $b_\delta^2 P_{\delta\delta}(k)$ (blue, dotted line),
$2 \fnl b_\phi b_\delta P_{\phi\delta}(k)$ (green, short-dashed),
and $\fnl^2 b_\phi^2 P_{\phi\phi}(k)$ (red, long-dashed line).  
The arrow indicates the approximate 
minimum $k$ probed by an all-sky survey out to $z=1$. 
}
\label{figP}
\end{figure}

Inevitably, we now also obtain a contribution to
the power spectrum $c_\phi^2 \fnl^2 P_{\phi \phi}\left(k\right)$, i.e., the 
linear theory power spectrum is now 
\begin{equation}
P_{gg}\left(k\right) = c_\delta^2 P_{\delta_L \delta_L}\left(k\right)+
2 c_\delta c_\phi \fnl P_{\phi \delta_L}\left(k\right)+
c_\phi^2 \fnl^2 P_{\phi \phi}\left(k\right)+...
\label{eqlinPwphi}
\end{equation} 
Note that the decision to write $c_\phi \fnl$ instead of including the 
$\fnl$ factor in $c_\phi$ is suggestive but ultimately arbitrary, since 
$c_\phi$ can in principle depend on $\fnl$.  The divergent
correction to $c_\phi$ is of generally the same order of magnitude, i.e., 
proportional to
$\sigma_{\delta\delta}$, as the correction to $c_\delta$ introduced
by \cite{2006PhRvD..74j3512M}, which gives us some reason to expect $b_\delta$
and $b_\phi$ to be of similar order (I use $b$ in place of $c$ to indicate
the renormalized parameter).
Figure \ref{figP} shows an example of this power spectrum.

Equation (\ref{eqlinPwphi}) is really the bottom line result of the paper, 
but for
completeness I compute all the higher-order terms arising from the new term
\begin{eqnarray}
\label{eqnewnonlinearphiterms}
2 c_\phi \fnl \left[\frac{c_{\delta^2}}{2}P_{\phi \delta^2}\left(k\right)+
\frac{c_{\delta^3}}{6}P_{\phi \delta^3}\left(k\right)\right] &=&
4 c_\phi c_{\delta^2}\fnl^2 
P_{\phi \phi}\left(k\right)
\int \frac{d^3 \vq}{\left(2\pi\right)^3} P_{\phi \delta_L}\left(q\right)
M\left(\left|\vk-\vq\right|\right)  \\ \nonumber 
&+&c_\phi\fnl \left(\frac{68}{21}c_{\delta^2}+ c_{\delta^3}\right)
P_{\phi \delta_L}\left(k\right) \sigma_{\delta\delta}^2~.
\end{eqnarray}
Note that each piece here has a near-divergent part, however, each of these 
is neatly canceled by the renormalizations we are already doing.
The first term has the same form as the ones we dealt with in 
Eq. (\ref{eqnontrivialbias}), so one can see that the redefinition
$b_\phi = c_\phi + 2 c_{\delta^2} \sigma_{\delta\delta}^2$ will cancel
the divergent part of the first term in Eq. (\ref{eqnewnonlinearphiterms}). 
Similarly, the original renormalization of $c_\delta$ in 
\cite{2006PhRvD..74j3512M}, $b_\delta=c_\delta+
\left(\frac{34}{21}c_{\delta^2}+\frac{c_{\delta^3}}{2}\right)
\sigma_{\delta\delta}^2$,
cancels the rest of Eq. (\ref{eqnewnonlinearphiterms}), when 
$b_\delta \delta$ is cross multiplied with $c_\phi \fnl \phi$. 
Things work out so neatly simply because
$P_{\phi X}\left(k\right)=M^{-1}\left(k\right)P_{\delta X}\left(k\right)$
(where $X$ could be anything), so there is no nontrivial new behavior when 
calculating cross terms with $\phi$ instead of $\delta$.
The cross terms between the new linear $\phi$ term and higher-order
gravitational corrections
to $\delta$ are negligible because the latter only contribute at small scales, 
while $\phi$ is only significant at large scales (note, however, that if one
were trying to exploit the very small non-Gaussian signal that is present at 
relatively high $k$, these terms might come into play).

This linear-quadratic ($\delta \delta^2$) cross term is the only new 
interesting one because
the linear-linear term is of course just the usual linear bias times the 
nonlinear power spectrum (shown to be unaffected by non-Gaussianity in the
previous section), while the quadratic-quadratic and linear-cubic 
terms are already 4th order in $\phi$, without the non-Gaussian part.

After the divergent pieces are subtracted from Eqs. (\ref{eqnontrivialbias})
and (\ref{eqnewnonlinearphiterms}), the $k$-dependent residuals are tiny for 
any reasonable $\fnl$, so I drop them.  
The final power spectrum is thus
\begin{eqnarray}
P_{gg}\left(k\right) &=& b_\delta^2 P_{\delta \delta}\left(k\right)+
2 b_\delta b_\phi \fnl P_{\phi \delta}\left(k\right)+
b_\phi^2 \fnl^2 P_{\phi \phi}\left(k\right) \\ \nonumber &+&
\frac{b_{\delta^2}^2}{2}\int \frac{d^3\vq}{\left(2 \pi\right)^3}
P_{\delta\delta}\left(q\right)
\left[P_{\delta \delta}\left(\left|\vk-\vq\right|\right)-
P_{\delta\delta}\left(q\right)\right] \\ \nonumber &+&
2~b_{\delta^2} \int \frac{d^3\vq}{\left(2 \pi\right)^3} 
P_{\delta\delta}\left(q\right)
P_{\delta\delta}\left(\left|\vk-\vq\right|\right)
J^{\left(2\right)}_S\left(\vq,\vk-\vq\right) \\ \nonumber 
&+& N~,
\end{eqnarray}
where the last three terms are those found in \cite{2006PhRvD..74j3512M}, 
unrelated to primordial non-Gaussianity.

\section{Conclusions and Discussion \label{secconclusions}}

In Sec. \ref{secmassdensity}, we found that the corrections to the mass power
spectrum arising from the non-Gaussian initial conditions are naively infinite,
but the divergence can be removed by a renormalization of the amplitude of 
perturbations, after which
there is no significant effect of non-Gaussianity for realistic models.  
It is interesting to note that a sufficiently large numerical simulation would
have to deal with this issue explicitly, because the most natural method of
implementing initial conditions effectively implements the unrenormalized 
version \cite{2008PhRvD..77l3514D}; however, the dynamical range of typical
simulations is small enough that this effect is unlikely to be noticeable (the 
box size and finite resolution 
provide large and small-scale cutoffs).
The issues discussed in Sec. \ref{secmassdensity} should be revisited when 
considering other forms of non-Gaussianity. 

In Sec. \ref{secbias} I computed the power spectrum of biased tracers of 
mass density and found a large higher-order perturbative correction that 
can only be removed by renormalization if we add a term to the 
standard linear bias model, proportional to $\phi$, producing the model 
\begin{equation}
\delta_g=
c_\delta~\delta +c_\phi \fnl \phi+
\frac{1}{2}~ c_{\delta^2}~\left( \delta^2-\sigma^2\right) +
\frac{1}{3!}~ c_{\delta^3}~ \delta^3 +\epsilon+\orderfour~.
\end{equation}
This new term will propagate into the computation of all statistics, not just
the power spectrum.  Note that $\phi$ should not be confused with the Newtonian
potential at the time of observation -- it is defined in terms of $\delta$ 
by Eq. (\ref{eqPoisson}).

One might ask at this point: ``once there is a linear term, why did we not add 
higher-order terms in $\phi$?''
The short answer is simply that we did not need to.  We added the minimal term
necessary to produce a well-behaved calculation.  A calculation to even higher
order would no doubt require the addition of higher order counter terms, and 
guide their form.  It seems likely that the higher-order terms would be 
completely negligible, like
the corrections to the mass power spectrum, but, as we have learned, one should
really compute them instead of assuming. 

Measuring $\fnl$ accurately depends on
predicting $b_\phi$ for an observable in just the same way as measuring the 
amplitude of the matter power spectrum depends on a prediction of the usual 
bias $b_\delta$. 
By comparison to the results of \cite{2008PhRvD..77l3514D,
2008ApJ...677L..77M,2008JCAP...08..031S}, we can determine 
the value of $b_\phi$ predicted by the
halo model.  As emphasized by \cite{2008JCAP...08..031S}, however, it should be
kept in mind that this kind of prediction makes some assumptions that are not
guaranteed to be true, e.g., the galaxy population in a halo
might depend on something other than halo mass alone, like merging history, 
which affects the clustering of the population. 
Halo model predictions will give a good estimate for the size of effect we
expect for a given $\fnl$ and type of galaxy, but if we are fortunate enough to 
make a detection it will be very 
difficult to measure $\fnl$ to high precision using only this power spectrum 
effect.  With that caveat, Eq. (18) of \cite{2008JCAP...08..031S} leads to 
\begin{equation}
b_\phi = 2 \left(b_\delta-1\right) \delta_c=3.372 \left(b_\delta-1\right)~.
\label{halobphi}
\end{equation}

\begin{figure}
\resizebox{\textwidth}{!}{\includegraphics{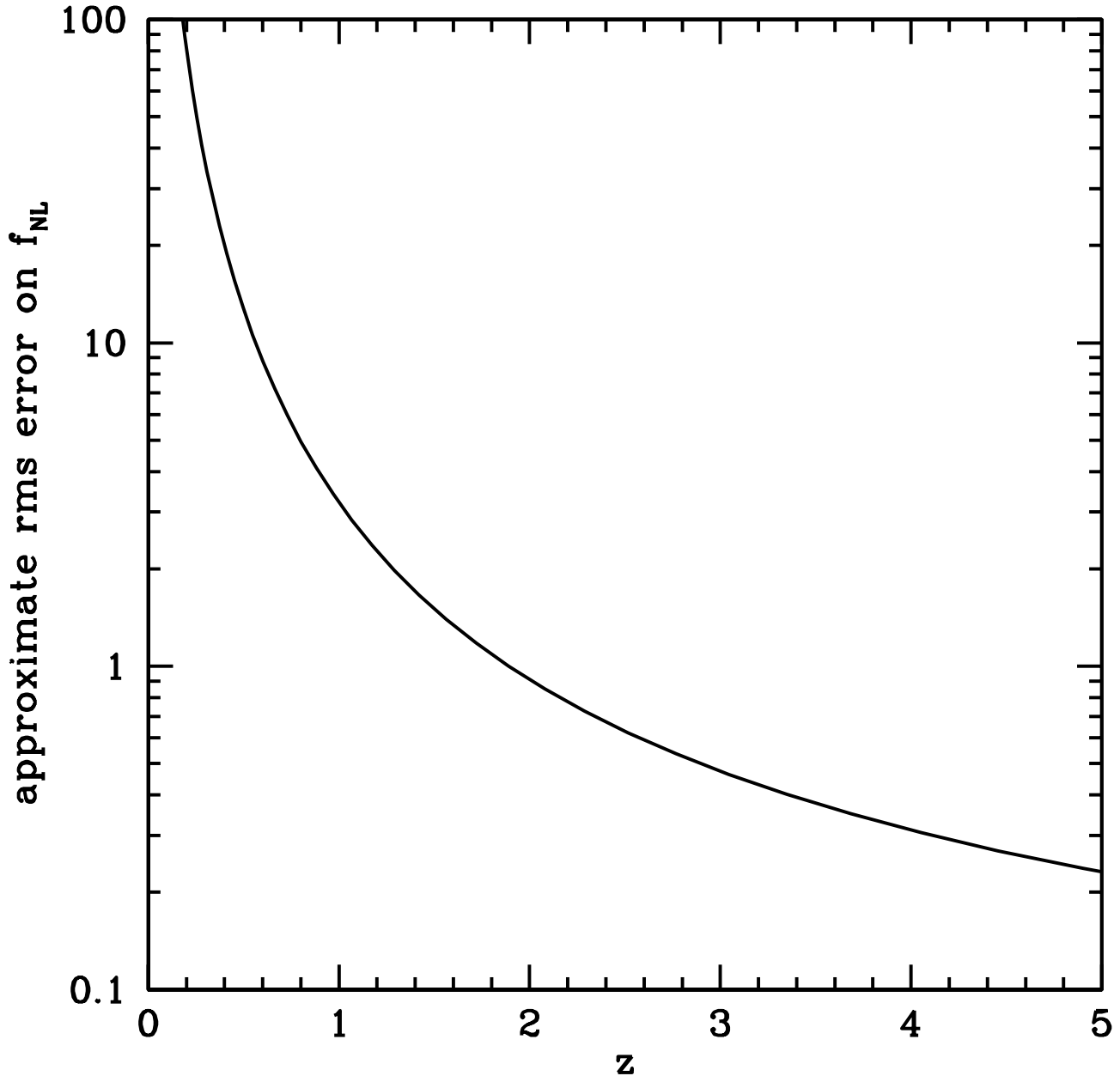}}
\caption{
Rms error on $\fnl$ for a well-sampled all-sky survey out to $z$, ignoring 
redshift-space, geometric, and evolutionary distortions.  Rms detection 
significance is essentially linear in $\fnl$ (out to a $\sim 5 \sigma$
level of detection), and scales roughly as ${\rm volume}^{2/3}$ (as long as the
volume remains compact).
}
\label{figfnlcon}
\end{figure}

Reference \cite{2008PhRvD..77l3514D} computed that a very modest future 
survey, 
extending only to $z=0.7$ (while the bulk of the volume of the universe is at
higher redshift) can constrain $\fnl$ to $\sim \pm 10$.  
It is interesting to know at 
least roughly how well a larger survey can do.
For a well-sampled survey (i.e., no shot noise), the signal to noise of a
single mode is 
\begin{equation}
\frac{S}{N}_{\rm mode}=\frac{\left[1+
\left(\frac{k_\star}{k}\right)^2\right]^2-1}
{\left[1+\left(\frac{k_\star}{k}\right)^2\right]^2} ~,
\end{equation}
(note that a null test would not include the signal in the denominator, but 
this makes very little difference, as I will show) where 
\begin{equation}
k_\star=\left(\frac{3~b_\phi \fnl \Omega_{m,0}}
{2~ b_\delta D\left(a\right)}\right)^{1/2} \frac{h~ {\rm Mpc}^{-1}}{3000}
\end{equation}
I am assuming $T\left(k\right)\simeq 1$ on the relevant scales, as discussed
below.
I will ignore redshift space, geometric, and evolutionary distortions, and any 
confounding systematic errors or parameter degeneracies.  
The total signal to noise of a survey is approximately
\begin{equation}
\left(\frac{S}{N}\right)^2\simeq
\frac{V}{4 \pi^2} \int_{k_{\rm min}}^{k_{\rm max}} dk k^2 \left[\frac{S}{N}
\left(k\right)\right]^2
\end{equation}
where $k_{\rm min}\sim 2 \pi / V^{1/3}$ (the detection limit is not very 
sensitive to $k_{\rm min}$, as long as $k_{\rm min}$ is basically set by the 
survey volume). 
$k_{\rm max}$ should be something less than $\infty$, but, for interesting 
$k_\star$ the  
integral converges within the linear regime (the integral would
not converge if the $k$ dependence of the transfer function was included, 
however, the information gained this way would come from a tiny change in 
signal on scales where the signal would not be uniquely distinguishable from
changes in other parameters, and linearity cannot be safely assumed, so a 
projection including it would be unreliable).
The bulk of the signal comes from the range $2 k_\star \lesssim k \lesssim 
10 k_\star$.  Consequently, the bulk of the signal comes from the 
$\fnl \phi~\delta$ cross term, rather than the $\fnl^2 \phi~\phi$ term 
(cross correlating with an unbiased tracer is worse than autocorrelation, 
but essentially only because of the factor of 2 that is lost in the 
cross term).  
To a very good approximation,
$S/N=1$ when $k_\star \sim \pi V^{-1/3} \sim k_{\rm min}/2$ 
(e.g., this would be 
$\sim 0.8~\pi V^{-1/3}$ if I used the null test $S/N$ per mode and set 
$k_{\rm min}$ based on the diameter of a sphere instead of the side length of a 
cube, which makes it slightly smaller), which gives 
$\fnl b_\phi/ b_\delta D(a)\simeq 200 \left({\rm Gpc}/h\right)^2/V^{2/3}$
(for $\Omega_m=0.3$).  Assuming Eq. (\ref{halobphi}) and 
$b_\delta D(a)=2$, we have finally that $S/N=1$ when 
$\fnl \sim 125 \left[\left({\rm Gpc}/h\right)^2/V^{2/3}\right]  / 
\left[2/D(a)-1\right]$.  
I plot the value of this function for a survey 
covering all volume up to $z$ in Fig. \ref{figfnlcon}.
Note that the assumption $b_\delta D(a)=2$ appears to make $b_\delta$ 
unreasonably large at high redshift, however, the value of $b_\delta$ 
cancels in $b_\phi/b_\delta$ if $b_\delta$ is very large, so the result is not 
actually very sensitive to this assumption.
We see that there is plenty of volume in the Universe to measure $\fnl$ even
if it is less than 1, although this will of course take a huge amount of work. 
Reference \cite{2008arXiv0806.1046A} found similar results, using a variety of 
slightly
different assumptions, which testifies to the robustness of the projection 
(which ultimately just amounts to the statement that the survey should be
large enough to resolve the $k$ where the non-Gaussian-induced power becomes
roughly equal to the Gaussian power).  
Twenty-one cm intensity mapping 
\cite{2008MNRAS.383.1195W,2008PhRvL.100i1303C} may be a 
route to surveying all of the post-reionization volume.  

Even if one's tracer is weakly biased, so the halo
model predicts little signal, it may be possible to make a nonlinear 
transformation of the density field to enhance the bias; however, it is not 
clear that any such transformation
would be better than simply measuring the bispectrum.  It 
is interesting to note that
the power spectrum measurement proposed by  \cite{2008PhRvD..77l3514D} and 
discussed in this paper can be seen as nature's implementation of the
poor-person's bispectrum measurement suggested by \cite{2006PhRvL..97z1301C}, 
i.e., the cross-correlation of the field with the square of the field, where
here the squaring is done for us, as part of the formation of biased
structure. The significance of the difference between the natural and artificial
versions should not be underestimated -- exploiting the natural version we can 
observe
the effect on large scales even if limited resolution or noise prevents us from 
probing the small scales that would be required to perform a similar 
transformation artificially.
\cite{2006PhRvL..97z1301C,2008PhRvD..77j3506C} found that $\fnl$ may be 
measurable to $\sim 0.01$ using the bispectrum of high redshift 21cm 
observations and probing down
to very small scales.  In this paper I have excluded the possibility of 
measuring the signal using small scales
(by assuming the transfer function
is 1 in the signal-to-noise ratio calculation), but it may be useful, 
especially at high 
redshift, to explore the 
possibility of using smaller scales and combining the bispectrum and power 
spectrum.  

I thank Niayesh Afshordi and Latham Boyle for helpful discussions.

\bibliography{cosmo,cosmo_preprints}

\end{document}